\documentclass[journal,a4paper]{IEEEtran}
\usepackage[utf8]{inputenc}
\usepackage{amsmath}
\usepackage{mathtools}
\usepackage{graphicx}
\usepackage{float}
\usepackage{pgfplots}
\usepackage{tikz}
\usetikzlibrary{plotmarks}
\usetikzlibrary{external}
\usetikzlibrary{arrows}
\usetikzlibrary{shapes,decorations}
\usetikzlibrary{colorbrewer}
\usepackage[font=footnotesize]{caption}
\usepackage{cite}
\pgfplotsset{compat=1.14}
\usepackage{xcolor}
\usepackage{hyperref}
\usetikzlibrary{shapes}
\usepackage{pgfplotstable}
\usetikzlibrary{plotmarks}
\usetikzlibrary{external}
\usepackage{mathtools}
\DeclarePairedDelimiter{\ceil}{\lceil}{\rceil}
\DeclarePairedDelimiter{\floor}{\lfloor}{\rfloor}
\usetikzlibrary{arrows}
\usetikzlibrary{shapes,decorations}
\usetikzlibrary{colorbrewer}
\usepackage{caption}
\pgfplotsset{compat=1.14}
\usepackage[utf8]{inputenc}
\newcommand{\rom}[1]{\uppercase\expandafter{\romannumeral #1\relax}}

\usepackage{amssymb}

\begin{document}

\title{A Modulation Format Correction Formula for the Gaussian Noise Model in the Presence of Inter-Channel Stimulated Raman Scattering}

\author{Daniel~Semrau,~\IEEEmembership{Student Member,~IEEE,}
Eric~Sillekens,~\IEEEmembership{Student Member,~IEEE,}
Robert~I.~Killey,~\IEEEmembership{Senior Member,~IEEE,} and~Polina~Bayvel,~\IEEEmembership{Fellow,~IEEE,~Fellow,~OSA}
\thanks{This work was supported by a UK EPSRC programme grant TRANSNET (EP/R035342/1) and a Doctoral Training Partnership (DTP) studentship for Daniel Semrau.}
\thanks{D. Semrau, E. Sillekens, Robert I. Killey and P. Bayvel are with the Optical Networks Group, University College London, London
WC1E 7JE, U.K. (e-mail: \{uceedfs; e.sillekens; r.killey; p.bayvel\}@ucl.ac.uk.)}
}

\maketitle

\markboth{\today}%
{}

\begin{abstract}
A closed-form formula is derived, which corrects for the modulation format dependence of the Gaussian Noise (GN) model in the presence of inter-channel stimulated Raman scattering (ISRS). The analytical result enables a rapid estimate of the nonlinear interference (NLI) for arbitrary modulation formats and avoids the need for complex integral evaluations and split-step simulations. It is shown that the modulation format dependent NLI can be approximated by two contributions, one originating from a single span and one asymptotic contribution for a large number of spans. The asymptotic contribution is solved in closed-form for an arbitrary link function, making the result applicable for generic fiber systems using lumped, distributed or hybrid amplification schemes. The methodology is applied to the ISRS GN model and a modulation format correction formula in closed-form is derived which accounts for an arbitrary number of spans, inter-channel stimulated Raman scattering, arbitrary launch power distributions and wavelength dependent dispersion and attenuation. The proposed formula is validated by numerical simulations over the entire C+L band for multiple fiber types. 
\end{abstract}


\IEEEpeerreviewmaketitle

\section{Introduction}
\IEEEPARstart{A}{nalytical} models experienced substantial popularity over the last decade, as they provide a low complexity alternative to split-step simulations with good agreement to experiments \cite{Nespola_2014_gvo,Cai_2014_tpo,Nespola_2015_evo,Galdino_2016_edo,Saavedra_2017_eio,Saavedra_2017_eao,Saavedra_2018_isr}. Perturbation models that estimate nonlinear interference (NLI) are key for rapid and efficient system design \cite{Hasegawa_2017_ofd}, achievable rate estimations of point-to-point links \cite{Semrau_2016_air,Bosco_2011_aro,Shevchenko_2016_air} and physical layer aware network optimization \cite{ Ramamurthy_1999_iot, Anagnostopoulos_2007_pli}.
\par 
\ 
The Gaussian Noise (GN) model \cite{Splett_1993_utc,Tang_2002_tcc, Poggiolini_2012_tgm} is extensively applied throughout research and industry, as it offers very low computational complexity while being reasonably accurate for high cardinality modulation formats. Additionally, the GN model offers approximations in closed-form for a variety of optical transmission scenarios \cite{Splett_1993_utc,Bosco_2011_aro,Louchet_2003_amf,Chen_2010_cef,Poggiolini_2012_tgm,Savory_2013_aft,Johannisson_2014_mon,Poggiolini_2015_asa, Semrau_2017_ace}. Such closed-form approximations enable performance estimations within picoseconds and are vital for real-time applications and on-the-fly optimizations. 
\par 
\ 
\begin{figure}
    \includegraphics[]{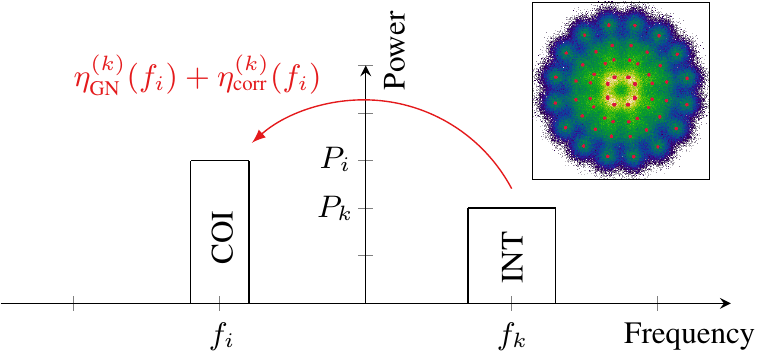}
\caption{Illustration of the NLI contribution of the Gaussian Noise (GN) model and the modulation format correction on the channel of interest (COI) originating from a single interferer (INT). The channels exhibit arbitrary modulation formats, power levels, bandwidths and center frequencies. The shown modulation format is taken from \cite{Chen_2018_g6c}.}
\label{fig:xpm_scheme}
\end{figure}
The major source of inaccuracy of the GN model stems from its signal Gaussianity assumption, which assumes that the signal can be written as a Gaussian process at the transmitter. As a consequence the model significantly overestimates the NLI for low cardinality modulation formats and in optical links with low accumulated dispersion. More complex models, in integral form, have been introduced to account for non-Gaussian modulation formats \cite{Mecozzi_2012_nsl, Secondini_2012_afc, Dar_2013_pon, Dar_2014_aon, Jiang_2014_tgm, Carena_2014_emo, Ghazisaeidi_2017_ato}. However, for those models to be applicable in a real-time environment and to avoid the high computational complexity of modulation format aware approaches, approximations in closed-form are necessary. A closed-form modulation format correction formula in the asymptotic limit of a large span number and optical bandwidths limited to C-band has been derived in \cite{Poggiolini_2015_asa}.
\par 
\ 
All of the aforementioned modeling approaches do not account for inter-channel stimulated Raman scattering (ISRS) and are therefore not applicable to ultra-wideband transmission systems that operate at optical bandwidths beyond C-band (5 THz). Extensions of the conventional GN model to account for ISRS, named ISRS GN model, have been recently proposed in integral form \cite{Semrau_17_ard, Roberts_2018_co, Cantono_2018_mti, Semrau_2018_tgn, Cantono_2018_oti, Semrau_2018_tig}. An approximation in closed-form has been first proposed in  \cite{Semrau_17_ard} which recently has been significantly generalized to include arbitrary launch power distributions, wavelength dependent dispersion and a more accurate description of ISRS \cite{Semrau_2018_aca, Semrau_2019_aca}. A comparible formula, based on the same approach and with a similar result, has been reported in \cite{Poggiolini_2018_agg}. 
\par 
\ 
However, the works \cite{Semrau_17_ard, Roberts_2018_co, Cantono_2018_mti, Semrau_2018_tgn, Cantono_2018_oti, Semrau_2018_tig, Semrau_2018_aca,Semrau_2019_aca, Poggiolini_2018_agg} assume signal Gaussianity and the impact of the modulation format has not been addressed analytically in the context of ultra-wideband transmission systems, where ISRS cannot be neglected. More importantly, a closed-form formula suitable for real-time and yet accurate performance estimations of arbitrary modulation formats has not been reported to date.
\par 
\
In this paper, a modulation format correction formula for the ISRS GN model is derived. Is is shown that the modulation format correction can be, in general, approximated by two contributions. One contribution that accounts for the modulation format correction after one span and one accounting for the correction in the asymptotic limit of a large span number. The asymptotic contribution is solved for an arbitrary transmission system, described by a link function, yielding a generic asymptotic modulation format correction formula. Both contributions, yielding a formula for any number of spans, are solved particularly for the ISRS GN model. The proposed formula accounts for arbitrary launch power distributions, arbitrary number of spans, arbitrary modulation formats, lumped amplification and ultra-wideband effects such as inter-channel stimulated Raman scattering and wavelength dependent dispersion and attenuation. The finding is validated by numerical simulations for a low dispersive non-zero dispersion-shifted fiber (NZDSF) and a high dispersive standard single mode fiber (SMF) over 10 THz optical bandwidth. 
\par 
\ 
The paper is organized as follows: In Section \ref{se:Modulation_Format_Correction}, the  formalism to analytically describe the impact of the modulation format on the NLI is reviewed and the assumptions necessary to solve it in closed-form are described. The methodology is then applied to the ISRS GN model in Section \ref{sec:Modulation_format_correction_formula_for_the_ISRS_GN_model} to obtain a modulation format correction formula in closed-form, the key result of this work. The formula is validated by numerical simulations in Sections \ref{sec:numerical_validation} and \ref{sec:full_performance_prediction}.

\section{Modulation Format Correction}
\label{se:Modulation_Format_Correction}
After coherent detection, electronic dispersion compensation and neglecting the impact of transceiver noise, the signal-to-noise ratio (SNR) of the channel of interest (COI) $i$ can be calculated as
\begin{equation}
\begin{split}
\label{eq:SNR}
\text{SNR}_i \approx \frac{P_i}{P_\text{ASE} + \eta_n P_i^3},
\end{split}
\end{equation}
where $P_i$ is the launch power of channel $i$ at the transmitter and $P_\mathrm{ASE}$ is the accumulated amplified spontaneous emission (ASE) noise originating from optical amplifiers. ASE noise inflation as a result of gain equalization can be included by a channel dependent $P_\text{ASE}$. The nonlinear interference coefficient $\eta_n\left(f_i\right)$ after $n$ fiber spans is dependent on the center frequency $f_i$ of the COI. 
\par 
\ 
Although modulation format independent models, such as the GN model, are in relative good agreement with experiments using high cardinality modulation formats, the NLI has been shown to be dependent on the transmitted modulation format. This dependence is particularly high for low cardinality modulation formats and short to medium-haul transmission distances. Substantially more complex models have been introduced in order to accurately model the impact of arbitrary modulation formats \cite{Mecozzi_2012_nsl, Dar_2013_pon, Dar_2014_aon, Secondini_2012_afc, Jiang_2014_tgm, Carena_2014_emo, Ghazisaeidi_2017_ato}. Most results show that the total NLI can be separated into two contributions, one 'GN-like' modulation format independent contribution and a correction term that accounts for the transmitted modulation format. The NLI coefficient can hence be written as 
\begin{equation}
\begin{split}
\eta_n\left(f_i\right) = \eta_{\text{GN,}n}\left(f_i\right) + \eta_{\text{corr.},n}\left(f_i\right),
\label{eq:total_NLI_coeff}
\end{split}
\end{equation}
where $\eta_{\text{GN,}n}$ is the GN model contribution and $\eta_{\text{corr.},n}$ is the modulation format correction term. For Gaussian modulated signals the correction term $\eta_{\text{corr.},n}$ vanishes and one obtains the GN model description of NLI. 
\par 
\ 
In \cite{Jiang_2014_tgm, Carena_2014_emo}, a complete set of formulas in integral form has been derived to correct for arbitrary modulation formats for self-phase modulation (SPM/SCI), cross-phase modulation (XPM/XCI) and four-wave mixing (FWM/MCI) NLI contributions. In the approach taken in this work, only the most dominant contribution, the cross-phase modulation contribution, is taken into account. It has been shown that this approach is sufficiently accurate with only a minor loss in accuracy \cite{Mecozzi_2012_nsl, Secondini_2012_afc,Dar_2013_pon, Dar_2014_aon, Poggiolini_2015_asa}. The approach is particularly accurate in high dispersive fibers such as standard single mode fiber (SMF) and systems with high symbol rates or channel spacings. Two conditions that are prevailing in legacy systems and likely to be installed in next generation optical transmission systems.
\par 
\ 
The XPM assumption, taking only XPM-like terms into account, evaluates the NLI within the channel of interest originating from a single interferer (INT). The XPM assumption is illustrated in Fig. \ref{fig:xpm_scheme}. The total NLI is obtained by summing over all COI-INT pairs present in the transmitted signal. Mathematically, the total XPM modualtion format correction is
\begin{equation}
\begin{split}
&\eta_{\text{corr,}n}\left(f_i\right)=\sum_{k=1,k\neq i}^{N_\text{ch}}\eta_{\text{corr,}n}^{\left(k\right)}(f_i),
\label{eq:XPM_eta}
\end{split}
\end{equation}
where $\eta_{\text{corr,}n}^{\left(k\right)}(f_i)$ is the XPM contribution of a single interfering channel $k$ on channel $i$. 
\par 
\
The dominant XPM modulation format correction term is given by \cite[Eq. (17)]{Carena_2014_emo}
\begin{equation}
\begin{split}
&\eta_{\text{corr,}n}^{\left(k\right)}(f_i) = \frac{80}{81}\left(\frac{P_k}{P_i}\right)^2 \frac{\gamma^2\Phi}{B_k^3} \int_{-\frac{B_i}{2}}^{\frac{B_i}{2}} df_1   \\
&\cdot\left|\int_{-\frac{B_k}{2}}^{\frac{B_k}{2}} \mu\left(f_1+f_i,f_2+f_i+\Delta f,f_i\right)\sum_{m=0}^{n-1}e^{jmf_1\left(f_2+\Delta f\right)\phi}df_2\right|^2,
\label{eq:XPM_integral_corr}
\end{split}
\end{equation}
where $B_i$ and $B_k$ are the bandwidth of COI $i$ and INT $k$, respectively, $\Delta f = f_k-f_i$ is the channel frequency separation between COI and INT, $\gamma$ is the nonlinearity coefficient and $\phi=-4\pi^2\left[\beta_2+\pi\beta_3(f_i+f_k)\right]L$ is a phase mismatch term accounting for coherent NLI accumulation, with the group velocity dispersion (GVD) parameter $\beta_2$ and its linear slope $\beta_3$ at the reference wavelength. In \eqref{eq:XPM_integral_corr}, it is assumed that the variation of the wavelength dependent dispersion is negligible over a single channel bandwidth. $\mu\left(f_1,f_2,f_i\right)$ is the link function of a single span. The link function is addressed in more detail in the following sections.
\par 
\ 
$\Phi=\frac{\mathrm{E}\left[\left|X\right|^4\right]}{\mathrm{E}^2\left[\left|X\right|^2\right]}-2$ is the excess kurtosis of the transmitted modulation format. The excess kurtosis of a few selected modulation formats are listed in Table \ref{tab:kurtosis}. Most modulation formats exhibit a negative excess kurtosis resulting in a negative modulation format correction $\eta_{\text{corr,}n}^{\left(k\right)}(f_i)$ and reduced nonlinear interference as a consequence. Loosely spoken, the modulation format correction is smaller for more 'Gaussian-like' modulation formats. This property does not only extend to the coordinates of the individual symbols but also to their respective probabilty of occurance \cite{Fehenberger_2016_ops}. 
\begin{table}
\renewcommand{\arraystretch}{1.2}
\centering
\caption{Excess kurtosis of selected modulation formats.}
\label{tab:kurtosis}
  \begin{tabular}{ l | c }
   \textbf{Modulation format} & Excess kurtosis $\Phi$ \\  \hline
    uniform QPSK & -1\\ \hline
    uniform 16-QAM &-0.6800\\ \hline
    uniform 64-QAM & -0.6190\\ \hline
    uniform 256-QAM &-0.6050\\ \hline
    uniform $\infty$-QAM & -0.6000\\ \hline
    geom. shaped 64-QAM (12 dB SNR, GMI)\cite{Chen_2018_g6c} & -0.3403\\ \hline
        proba. shaped 64-QAM (12 dB SNR,  MI) \cite{Kschischang_1993_ons}& -0.1871\\ \hline 
    Gaussian modulation & 0\\ \hline
  \end{tabular}
\end{table}
\par
\
The aim of the remainder of this paper is to find an approximation in closed-form of the modulation format correction in integral form \eqref{eq:XPM_integral_corr}. This gives the advantage of increase prediction accuracy for non-Gaussian modulation formats without sacrificing computation complexity and execution time.
\subsection{Modulation format correction for generic transmission systems in closed-form}
\label{sec:generic_mod_format_correction}
In this section, an approximation of the solution of Eq. \eqref{eq:XPM_integral_corr} for a generic transmission system is derived. An optical fiber transmission system that is described by the nonlinear Schr\"odinger equation can be, to first-order, described by a link function $\mu\left(f_1,f_2,f_i\right)$. The link function describes the nonlinear perturbation (i.e. the NLI) of three interacting frequencies on the frequency $f_i$ after propagation. 
\par 
\ 
The approach in this work relies on two key assumptions. The first assumption is that the channel separation of two interfering channels (COI and INT) is much greater than the channel bandwidth, which allows to approximate $f_2+\Delta f \approx \Delta f$ in \eqref{eq:XPM_integral_corr}. Mathematically, this conincides with the zeroth-order solution of the inner integral running over $f_2$ in \eqref{eq:XPM_integral_corr}. A detailed assessment of this assumption was carried out in \cite[Appendix C]{Semrau_2019_aca} and negligible impact on the total NLI was found. Applying the assumption results in 
\begin{equation}
\begin{split}
\eta_{\text{corr,}n}^{\left(k\right)}(f_i) \approx & \ \tilde{\gamma}\int_{-\frac{B_i}{2}}^{\frac{B_i}{2}} df_1 \left|\mu\left(f_1+f_i,f_k,f_i\right)\right|^2  \\
&\cdot\left|1+\sum_{m=1}^{n-1}\mathrm{sinc}\left(m\phi f_1 \frac{B_k}{2}\right)e^{jm\phi f_1\Delta f} \right|^2,
\label{eq:corr_integral_1d}
\end{split}
\end{equation}
with $\tilde{\gamma}=\left(\frac{P_k}{P_i}\right)^2\frac{80}{81} \frac{\gamma^2\Phi}{B_k}$, which is introduced for notational brevity. 
\par 
\ 
 Eq. \eqref{eq:corr_integral_1d} does not seem to have an analytical solution. However, in order to find an approximate solution, we analyze its asymptotic behavior for a large number of fiber spans $n$. For a large number of spans, the oscilating terms in \eqref{eq:corr_integral_1d} can be approximated by a Dirac delta function $\delta\left(x\right)$ as
\begin{equation}
\begin{split}
& \lim_{n \to \infty} \left|1+\sum_{m=1}^{n-1}\mathrm{sinc}\left(m\phi f_1 \frac{B_k}{2}\right)e^{jm\phi f_1\Delta f} \right|^2 \\ 
\approx & 1+\delta\left(f_1\right)\lim_{n \to \infty} C_n,
\label{eq:asympotic_appx}
\end{split}
\end{equation}
with normalization coefficient
\begin{equation}
\begin{split}
& C_n = \int_{-\infty}^{\infty} df_1 \left|\sum_{m=1}^{n-1}\mathrm{sinc}\left(m\phi f_1 \frac{B_k}{2}\right)e^{jm\phi f_1\Delta f} \right|^2.
\end{split}
\end{equation}
The sinc function is defined as $\mathrm{sinc}\left(x\right)=\frac{\mathrm{sin}\left(x\right)}{x}$. The normalization coefficient $C_n$ is an immediate consequence of the property $\int_{-\infty}^{\infty}\delta\left(x\right)dx=1$ of the Dirac delta function. 
\par 
\ 
The approximation \eqref{eq:asympotic_appx} is the second key assumption in this work and it can be motivated as follows: For $f_2=0$, both sides of \eqref{eq:asympotic_appx} yield infinity, making the $1$ negligible and as a result it can be pulled out of the absolute square term. For $f_2 \neq 0$, the oscillating terms add mostly out-of-phase (deconstructively) and they are further damped by the $\frac{1}{f_1}$ decay in the $\mathrm{sinc}\left(x\right)$ function. As a consequence, the sum over the oscillating terms can be approximated by a Dirac delta function, the oscillating terms are negligible with respect to $1$ and the $1$ can be, again, pulled out of the absolute square.
\par
\
As a result of the asymptotic approximation \eqref{eq:asympotic_appx}, the NLI coefficient can be approximated by only two contributions. One modulation format correction originating in the first span and an asymptotic contribution originating in the limit of a large span number. The NLI coefficient can be thus written as
\begin{equation}
\begin{split}
&\eta_{\text{corr,}n}^{\left(k\right)}(f_i)\approx \underbrace{\eta_{\text{corr,}1}^{\left(k\right)}(f_i)}_{\text{1st span corr.}} + \underbrace{\tilde{n}\cdot\eta_{\text{corr,}a}^{\left(k\right)}(f_i)}_\text{asympt. corr.}
\label{eq:corr_contributions}
\end{split}
\end{equation}
with 
\begin{equation}
\begin{split}
&\eta_{\text{corr,}1}^{\left(k\right)}(f_i) = \tilde{\gamma} \int_{-\frac{B_i}{2}}^{\frac{B_i}{2}} df_1 \left|\mu\left(f_1+f_i,f_k,f_i\right)\right|^2 
\label{eq:cf_eta1_corr}
\end{split}
\end{equation}
\begin{equation}
\begin{split}
&\eta_{\text{corr,}a}^{\left(k\right)}(f_i) = \tilde{\gamma} \left|\mu\left(f_i,f_k,f_i\right)\right|^2 \cdot  \lim_{n \to \infty} \frac{\partial }{\partial n} C_n 
\label{eq:etaa_cf_general}
\end{split}
\end{equation}
and
\begin{equation}
\begin{split}
&\tilde{n} =  \begin{cases}
    0,& \text{if } n =  1\\
    n,              & \text{otherwise}
\end{cases}
\end{split}
\end{equation}
The asymptotic contribution $\eta_{\text{corr,}a}^{\left(k\right)}(f_i)$ is valid in the limit for a large number of spans. In order to calculate the modulation format correction for any number of spans, the asymptotic contribution is approximation by a Taylor series with respect to the number of spans and truncated to first-order. As the asymptotic contribution after one span must be zero, only its slope (in the asymptotic limit) must be calculated as in \eqref{eq:etaa_cf_general}. This can be done exactly and the necessary identity is derived in Appendix \ref{sec:appendix_proof}. 
\par 
\ 
Using the identity \eqref{eq:appendix_statement}, derived in Appendix \ref{sec:appendix_proof}, the asymptotic contribution can be written \textit{exactly} in closed-form as 
\begin{equation}
\begin{split}
&\eta_{\text{corr,}a}^{\left(k\right)}(f_i) =  \tilde{\gamma}\left|\mu\left(f_i,f_k,f_i\right)\right|^2\\
&\cdot  \frac{2\pi }{\left|\phi\right| B_k^2}\left[\left(2\Delta f-B_k\right)\mathrm{ln}\left(\frac{2\Delta f-B_k}{2\Delta f+B_k}\right)+2B_k\right] 
\label{eq:cf_etaa_corr}
\end{split}
\end{equation}
Eq. \eqref{eq:cf_etaa_corr} represents a generic modulation format correction formula, valid in the asymptotic limit of a large span number. Remarkably, the asymptotic contribution does not require any further integration and it, therefore, represents a closed-form correction formula for \textit{any} optical fiber transmission system described by a link function $\mu\left(f_1,f_2,f_i\right)$. Eq. \eqref{eq:cf_etaa_corr} can be universally applied and it is one of the main results in this paper. 
\par 
\ 
To further increase the accuracy for arbitrary span numbers, the integral, describing the modulation format correction contribution after a single span \eqref{eq:cf_eta1_corr}, must be solved. However, this integral resembles the mathematical structure of the GN model contribution $\eta_{\mathrm{GN,}n}\left(f_i\right)$ (cf. \cite[Eq. (14)]{Semrau_2019_aca}) In other words, if there exists a closed-form approximation for the GN model contribution after one span for an arbitrary transmission system, a modulation format correction formula in closed-form for any number of spans immediately follows with the results in this paper using \eqref{eq:corr_contributions}\eqref{eq:cf_eta1_corr}\eqref{eq:cf_etaa_corr}.

\subsection{Link function of the ISRS GN model}
In this section, the link function of the GN model in the presence of inter-channel stimulated Raman scattering is reviewed. Equations \eqref{eq:corr_contributions}\eqref{eq:cf_eta1_corr}\eqref{eq:cf_etaa_corr} are then applied to the ISRS GN model link function in order to derive a modulation format correction formula for arbitrary span numbers. The ISRS GN model represents an extension of the conventional GN model to account for ultra-wideband effects such as the power transfer between propagating frequencies due to ISRS. The generic link function of the ISRS GN model is given by \cite[Eq. (9)]{Semrau_2018_tgn}
\begin{equation}
\begin{split}
&\mu\left(f_1,f_2,f_i\right) = \int_0^L d\zeta \ \frac{P_{\text{tot}}e^{-\alpha \zeta-P_{\text{tot}}C_{\text{r}} L_{\text{eff}}\cdot(f_1+f_2-f_i)}}{\int G_{\text{Tx}}(\nu)e^{-P_{\text{tot}}C_{\text{r}} L_{\text{eff}}\nu} d\nu} \\
&\cdot e^{j\tilde{\phi}\left(f_1,f_2,f_i,\zeta\right)},
\label{eq:proper_link_function}
\end{split}
\end{equation}
where $P_\text{tot}$ is the total optical launch power, $C_r$ is the slope of the linear regression of the Raman gain function, $L_\text{eff}=\frac{1-e^{-\bar{\alpha}\zeta}}{\bar{\alpha}}$ and $\tilde{\phi}=-4\pi^2(f_1-f_i)(f_2-f_i)\left[\beta_2+\pi\beta_3(f_1+f_2)\right]\zeta$. Eq. \eqref{eq:proper_link_function} can be used to calculate the nonlinear perturbation on $f_i$ after one span for any arbitrary frequency triplet $\left(f_1,f_2,f_i\right)$. Eq. \eqref{eq:proper_link_function} accounts for all occuring nonlinear mixing products, namely self-phase (SPM/SCI), cross-phase (XPM/XCI) and four-wave mixing (FWM/MCI) products. However, the proposed formulas \eqref{eq:corr_contributions}\eqref{eq:cf_eta1_corr}\eqref{eq:cf_etaa_corr}, only correct for the dominant mixing products which are XPM products. This restricts the frequency triplets to the XPM domain which is $\left(f_1+f_i,f_k,f_i\right)$ with $f_1 \in \left[-\frac{B_i}{2},\frac{B_i}{2}\right]$. 
\par 
\ 
In our previous work \cite[Eq. (18)]{Semrau_2019_aca}, an approximation of \eqref{eq:proper_link_function} has been derived under the XPM assumption and a first-order description of ISRS. It was demonstrated that the ISRS GN model link function is well approximated by
\begin{equation}
\begin{split}
&\mu\left(f_1+f_i,f_k,f_i\right) \approx -\frac{1+\tilde{T}_k}{-\alpha +j\phi_{i,k}f_1} + \frac{\tilde{T}_k}{-A +j\phi_{i,k}f_1} 
\label{eq:appx_link_function}
\end{split}
\end{equation}
where $\tilde{T}_k  = -\frac{P_{\text{tot}}C_{\text{r}}}{\bar{\alpha}}f_k$, $\phi_{i,k}=-4\pi^2\left(f_k-f_i\right)\left[\beta_2+\pi\beta_3(f_i+f_k\right]$ and $A=\alpha+\bar{\alpha}$. If not specified otherwise, it holds that $\bar{\alpha}=\alpha$. The parameter $\bar{\alpha}$ can be used to apply the proposed closed-formula in more general cases. Such cases include improved accuracy for non-uniform (tilted) launch power distributions, wavelength dependent attenuation and even the extension of the formula beyond 15~THz i.e. beyond the triangular region of the Raman gain spectrum. This is done by reinterpreting $\alpha$, $\bar{\alpha}$ and $C_r$ as channel \textit{dependent} quantities. The parameters are then matched to reproduce the actual power profile of each channel and the proposed modulation format correction formula can be applied. The drawback of this strategy is larger complexity as the Raman equations must be solved numerically and additional regression operations are necessary in order to obtain the channel dependent $\alpha$, $\bar{\alpha}$ and $C_r$. 
\par 
\ 
In order to obtain the modulation format correction of the ISRS GN model in closed-form, the approximated link function \eqref{eq:appx_link_function} must be inserted in \eqref{eq:cf_eta1_corr} and \eqref{eq:cf_etaa_corr}. As mentioned in Section \ref{sec:generic_mod_format_correction}, the integral that needs to be executed in \eqref{eq:cf_eta1_corr} resembles the GN model contribution after one span. This integral has been solved in our previous work \cite{Semrau_2019_aca}. Hence, the modulation format correction for the ISRS GN model can be obtained using the results in this paper and the GN contribution derived in closed-form in \cite{Semrau_2019_aca}. 
\subsection{Modulation format correction for the ISRS GN model in closed-form}
\label{sec:Modulation_format_correction_formula_for_the_ISRS_GN_model}
Using the modulation correction formula for a generic system, derived in Section \ref{sec:generic_mod_format_correction}, combined with the approximated link function \eqref{eq:appx_link_function} of the ISRS GN model, yields a modulation format correction formula for the ISRS GN model in closed-form as
\begin{equation}
\begin{split}
&\eta_{\text{corr.,}n}\left(f_i\right) \approx \frac{80}{81}\Phi\sum_{k=1,k\neq i}^{N_\mathrm{ch}} \left(\frac{P_k}{P_i}\right)^2\frac{\gamma^2}{B_k}\left\{\frac{1}{\phi_{i,k}\bar{\alpha}\left(2\alpha+\bar{\alpha}\right)} \right.\\
&\cdot\left[\frac{T_k-\alpha^2}{\alpha}\mathrm{atan}\left(\frac{\phi_{i,k}B_i}{\alpha}\right) +\frac{A^2-T_k}{A}\ \mathrm{atan}\left(\frac{\phi_{i,k}B_i}{A}\right)\right]\\
&\left.+ \frac{2\pi \tilde{n} T_k }{\left|\phi\right| B_k^2\alpha^2 A^2}\left[\left(2\left|\Delta f\right|-B_k\right) \log\left(\frac{ 2\left|\Delta f\right| -B_k}{ 2\left|\Delta f\right| +B_k}\right)+ 2B_k\right]\right\},
\label{eq:cf_modcorr_ISRSGN}
\end{split}
\end{equation}
with $\Delta f = f_k-f_i$, $\phi_{i,k}=-2\pi^2\left(f_k-f_i\right)\left[\beta_2+\pi\beta_3(f_i+f_k\right]$\footnote{Please note that $\phi_{i,k}$ is different in Eq. \eqref{eq:appx_link_function} and in Eq. \eqref{eq:cf_modcorr_ISRSGN}\eqref{eq:full_eta_cf}. The variable is redefined to be consistent with prior work \cite[Appendix A]{Semrau_2019_aca}.} and $T_k = \left(\alpha+\bar{\alpha}-P_{\text{tot}}C_{\text{r}}f_k\right)^2$. The formula is applicable for lumped-amplified links for optical bandwidths of up to 15 THz, as the formula relies on the triangular Raman gain spectrum. For larger optical bandwidths, the variables $\alpha$, $\bar{\alpha}$ and $C_r$ can be matched to the actual power profile in the fiber and the formula can be applied (cf. Section \ref{sec:generic_mod_format_correction}). The former summand in \eqref{eq:cf_modcorr_ISRSGN} corrects for the modulation format within a single span, while the latter summand corrects the modulation format contribution across multiple spans as described by \eqref{eq:corr_contributions}. The sum in \eqref{eq:cf_modcorr_ISRSGN} corrects for all interfering channels within the transmitted WDM signal. 

\section{Numerical Validation}
\label{sec:numerical_validation}
In this section the proposed closed-form correction formula \eqref{eq:cf_modcorr_ISRSGN} is validated by numerical simulations over the entire C+L band, covering 10 THz optical bandwidth. The validation is carried out for two fiber types, one high dispersive standard single mode fiber (SMF) and one low dispersive non-zero dispersion-shifted fiber (NZDSF).
\subsection{Simulation Setup}
\begin{figure*}
   \centering
    \includegraphics[]{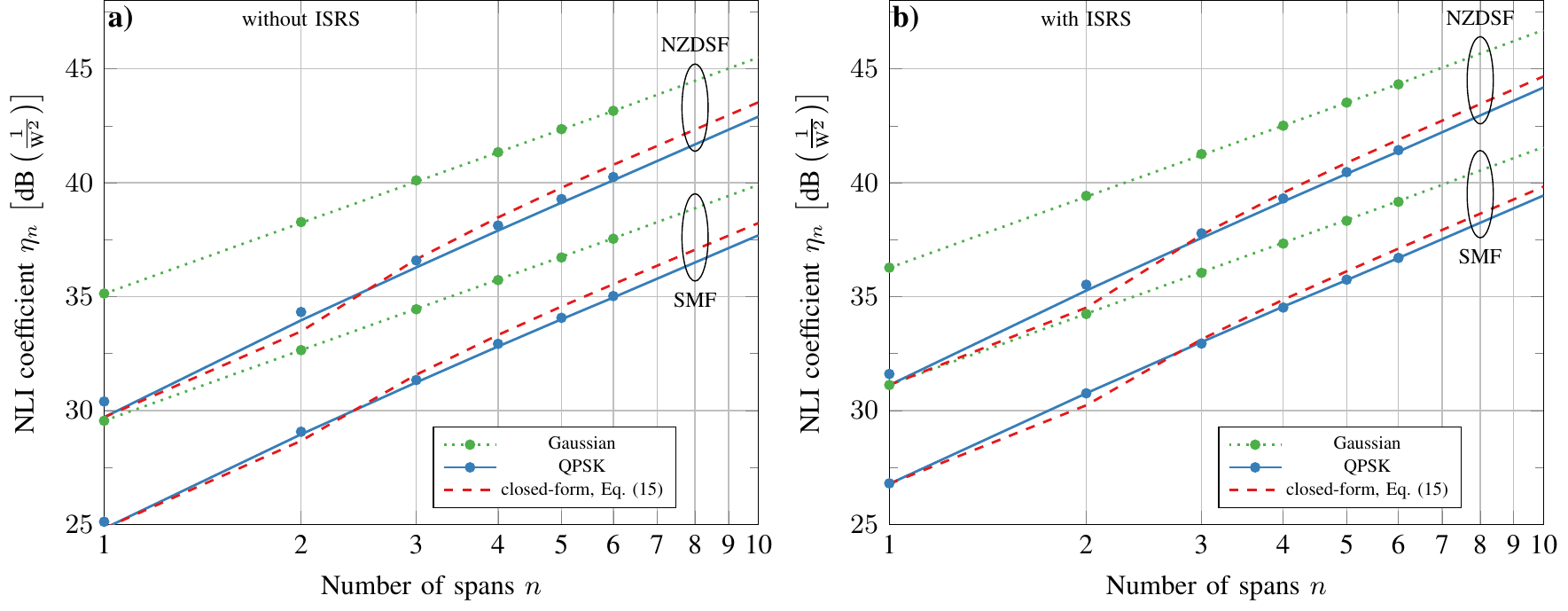}
\caption{The nonlinear interference coefficent as a function of fiber spans for the channel centered at $f_i = -4.0$ THz. The results were obtained by split-step simulations (markers) and the ISRS GN model \cite[Eq. (9)]{Semrau_2018_tgn} with the modulation format correction in integral form \eqref{eq:corr_integral_1d} (solid lines). The ISRS GN model with the modulation format correction in closed-form \eqref{eq:cf_modcorr_ISRSGN} is shown in dashed lines. In the case of of QPSK, only the channel of interest exhibits Gaussian modulation for validation purposes.}
\label{fig:NLI_coeff_-4THz}
\end{figure*}
\begin{figure*}
   \centering
    \includegraphics[]{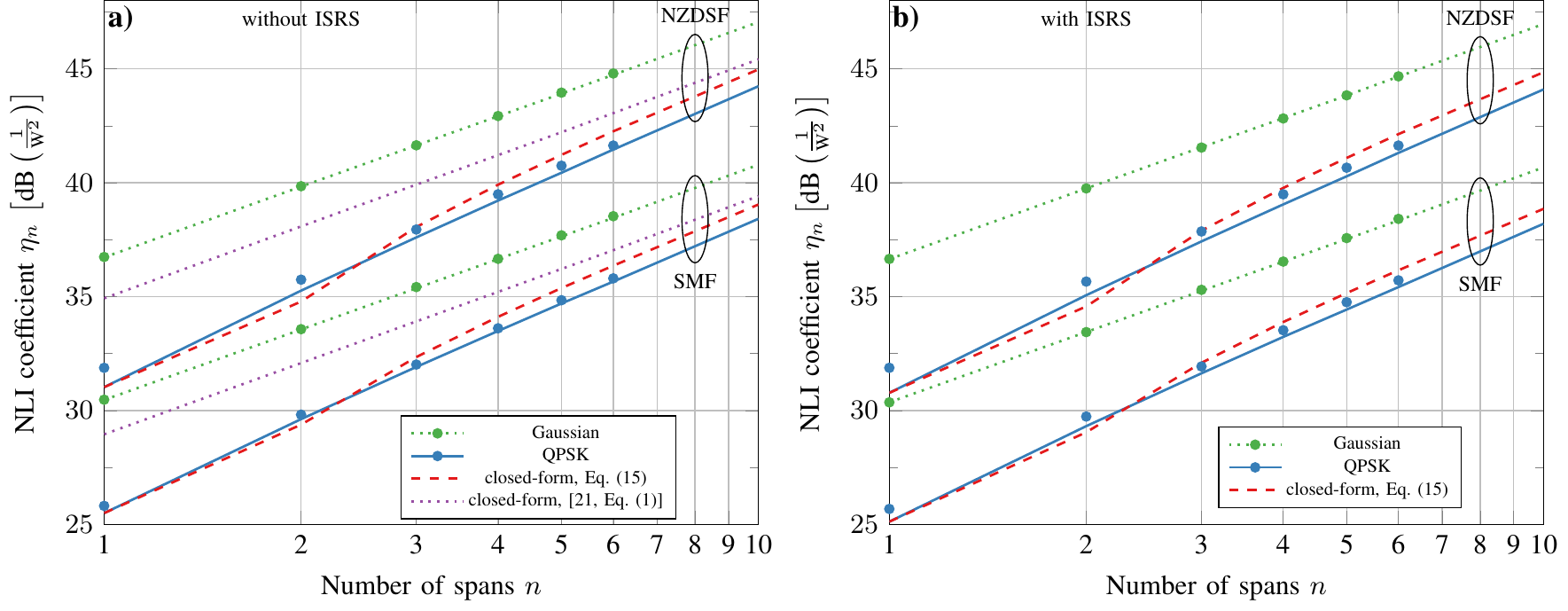}
\caption{The nonlinear interference coefficent as a function of fiber spans for the channel centered at $f_i = 0$ THz. The results were obtained by split-step simulations (markers) and the ISRS GN model \cite[Eq. (9)]{Semrau_2018_tgn} with the modulation format correction in integral form \eqref{eq:corr_integral_1d} (solid lines). The ISRS GN model with the modulation format correction in closed-form \eqref{eq:cf_modcorr_ISRSGN} is shown in dashed lines. In the case of of QPSK, only the channel of interest exhibits Gaussian modulation for validation purposes. For comparison the result of \cite[Eq. (1)]{Poggiolini_2015_asa} is shown, which proposed a modulation format correction formula in the absence of ISRS.}
\label{fig:NLI_coeff_0THz}
\end{figure*}
The validation was carried out by numerically solving the Manakov equation using the well established split-step Fourier method (SSFM). Inter-channel stimulated Raman scattering was included in the SSFM by applying a frequency dependent loss at every linear step, so that the signal power profile altered by ISRS, is obtained.
\par 
\    
A logarithmic step size distribution was implemented, where $0.25\cdot 10^6$ simulation steps were found to be sufficient for the considered launch powers and parameters of both fiber types. The launch powers were chosen to maximize the performance of the central channel assuming a 5~dB noise figure erbium-doped fiber amplifier (EDFA). 
\par 
\  
Gaussian symbols, drawn from a circular-symmetric Gaussian distribution and uniform QPSK symbols were used for transmission. In the case of QPSK modulation, the channels of interest, centered at $f_i=0$ THz and $f_i=-4.0$ THz, exhibit Gaussian modulation while the rest of the channels are modulated with QPSK symbols. This was done for validation purposes and only in the results of this section \ref{sec:numerical_validation}. This approach offers a more precise validation, as the proposed formula only corrects for modulation formats of interfering channels (XPM/XCI terms).
\par 
\  
The receiver consisted of digital dispersion compensation, ideal root-raised-cosine (RCC) matched filtering and constellation rotation. The SNR was ideally estimated as the ratio between the variance of the transmitted symbols $E[|X|^2]$ and the variance of the noise $\sigma^2$, where $\sigma^2=E[|X-Y|^2]$ and $Y$ represents the received symbols after digital signal processing. The nonlinear interference coefficient was then estimated via Eq. \eqref{eq:SNR}. In order to improve the simulation accuracy, four different data realizations were simulated and averaged for each transmission.
\par 
\  
Ideal, noiseless amplifiers were considered to ease the NLI computation and for a fair comparison between numerical simulation and modeling results. 
\begin{table}
\renewcommand{\arraystretch}{1.2}
\centering
\caption{System Parameters}
\label{tab:parameters}
  \begin{tabular}{ l | c c }
   \textbf{Parameters} & SMF & NZDSF \\  \hline
    Loss ($\alpha$) [dB/km]& 0.2 & 0.19\\ \hline
    Dispersion ($D$) [ps/nm/km]& 17.0 & 4.5\\ \hline
     Dispersion slope ($S$) [ps/$\text{nm}^2$/km]& 0.067 & 0.05\\ \hline
    NL coefficient ($\gamma$) [1/W/km]& 1.2 & 1.3\\ \hline
   Effective core area [$\mu\text{m}^2$]& 80 & 72\\ \hline
    Raman gain slope ($C_{\text{r}}$) [1/W/km/THz]&  0.028 & 0.031 \\ \hline
    Raman gain ($C_{\text{r}}\cdot 14$ THz) [1/W/km]& 0.39 & 0.44\\ \hline
    Channel Launch Power ($P_{i}$) [dBm]&   0 & -2\\ \hline
    Total Launch Power ($P_\text{tot}$) [dBm]&   24 & 22\\ \hline
     Symbol rate [GBd]&   \multicolumn{2}{c}{40}\\ \hline
    Channel Bandwidth ($B_{i}$) [GHz]&  \multicolumn{2}{c}{40.004} \\ \hline
    Channel spacing [GHz]&  \multicolumn{2}{c}{40.005} \\ \hline
    Number of channels &  \multicolumn{2}{c}{251} \\ \hline
    Optical bandwidth ($B_\text{tot}$) [THz]& \multicolumn{2}{c}{10.05} \\ \hline
    Reference Wavelength [nm]& \multicolumn{2}{c}{1550} \\ \hline
    Roll-off factor [\%]&  \multicolumn{2}{c}{0.01}  \\ \hline
    Number of symbols [$2^{x}$]& \multicolumn{2}{c}{17}  \\ \hline
    Simulation steps per span 
       
       [$10^6$]& \multicolumn{2}{c}{2.5}  \\ \hline
  \end{tabular}
\end{table}
\subsection{Results}
The nonlinear interference coefficient as a function of span numbers is shown in Fig. \ref{fig:NLI_coeff_-4THz} and \ref{fig:NLI_coeff_0THz} for the channels with center frequencies $f_i=-4.0$ THz and $f_i=0$ THz. The results are shown for both fiber types and for the case with ISRS and without ISRS. The (unphysical) case of not considering ISRS is shown for comparison. 
\par 
\ 
Markers represent simulation results, while lines represent modelling results. In the case of Gaussian modulation, the ISRS GN model in integral form \cite[Eq. (9)]{Semrau_2018_tgn}\cite[Eq. (2)]{Semrau_2018_tig} was used and is shown in dotted lines. In the case of QPSK modulation, the ISRS GN model in integral form was used with the modulation format correction in integral form \eqref{eq:XPM_integral_corr} (solid lines), as well as the modulation format correction formula in closed form \eqref{eq:cf_modcorr_ISRSGN} (dashed lines). 
\par 
\ 
The ISRS GN model has remarkable accuracy with numerical simulations exhibiting a negligible modelling error. In the case of QPSK modulation, the modulation format correction in integral form \eqref{eq:corr_integral_1d} models the impact of QPSK with good accuracy, despite one of the key assumptions made in this work ($\Delta f \gg \frac{B_k}{2}$). The average deviation between the modulation format correction in integral form and the numerical simulation is $0.26$ dB throughout the shown results. The error mostly stems from the XPM assumption and assumptions inherited by Eq. \eqref{eq:XPM_integral_corr}.The small impact of the assumption $\Delta f \gg \frac{B_k}{2}$ has been mathematically shown in \cite[Appendix C]{Semrau_2018_tgn} and is, therefore, not surprising.
\par 
\ 
\begin{figure*}
   \centering
    \includegraphics[]{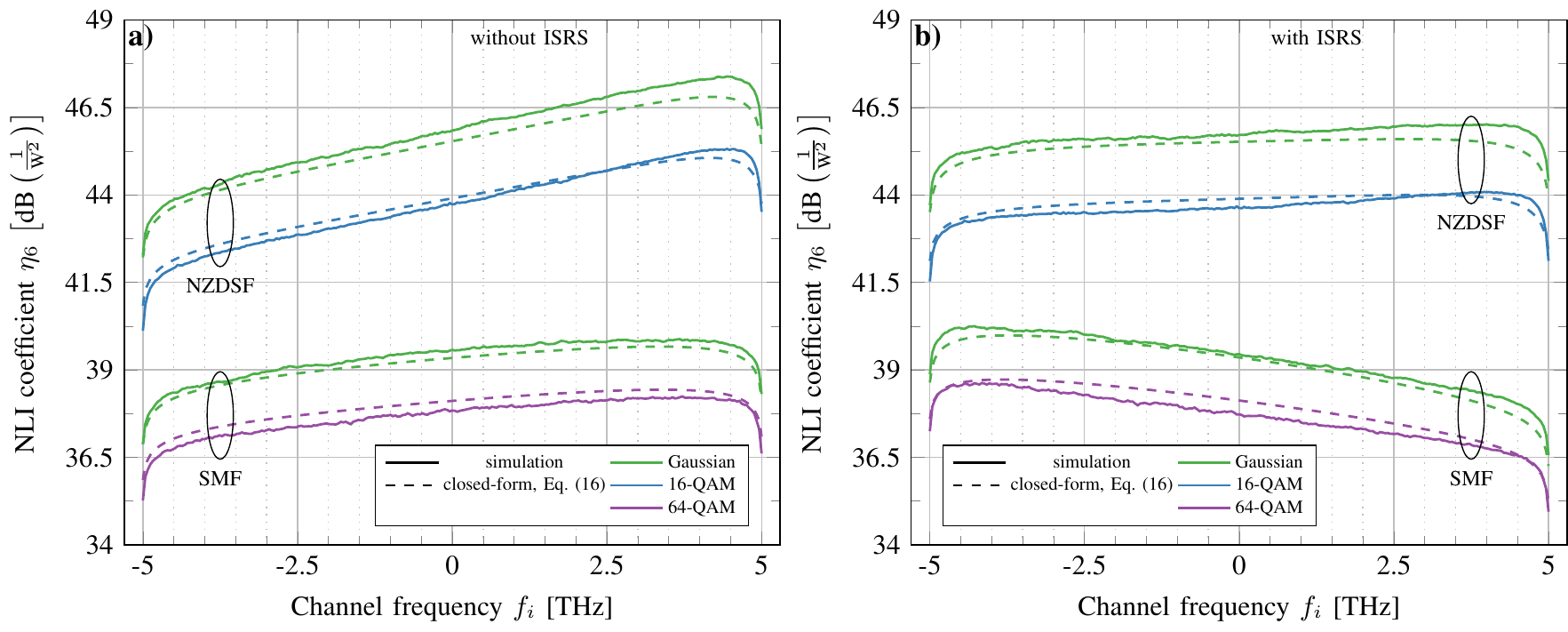}
\caption{The nonlinear interference coefficient as a function of channel frequency for different modulation formats and fiber types after 6x100 km spans. The results were obtained by split-step simulations and in closed-form using Eq. \eqref{eq:full_eta_cf}. Figures a) and b) show the case without and with considering inter-channel stimulated Raman scattering, respectively.}
\label{fig:cf_prediction}
\end{figure*}
The modulation format correction formula in closed-form \eqref{eq:cf_modcorr_ISRSGN} shows good accuracy throughout all number of spans. 
Throughout Fig. \ref{fig:NLI_coeff_-4THz} and \ref{fig:NLI_coeff_0THz}, the average absolute error is 0.45 dB between closed-form and numerical simulation. The majority of the mismatch can be traced back to the asymptotic assumption \eqref{eq:asympotic_appx} and its linear approximation of the asymptotic contribution \eqref{eq:etaa_cf_general} with respect to the span number (see Section \ref{sec:generic_mod_format_correction}). The mismatch is smaller for an increasing number of spans. Particularly, the error vanishes in the case of single span transmission and in the limit of infinte transmission spans. Not shown in the figures, the mismatch between modulation format correction in closed-form and integral form is $<0.1$ dB after 100 spans. 
\par 
\ 
The result in this paper is the first modulation format correction formula in the presence of ISRS. However, in the absence of ISRS (e.g. for optical bandwidths of at most 5 Thz), modulation format correction formulas are already available in the literature. To compare our results to previously published works in the absence of ISRS, \cite[Eq. (1)]{Poggiolini_2015_asa} is shown in Fig. \ref{fig:NLI_coeff_0THz}a), which proposes a modulation format correction formula in the absence of ISRS. As the result in \cite{Poggiolini_2015_asa} has been derived in the asymptotic limit of a large number of spans, it is rather inaccurate for the first few spans. The comparison shows that \eqref{eq:cf_modcorr_ISRSGN} is not only capable of correcting the modulation format in ISRS impaired systems but it also more accurate for than previously published results in the absence of ISRS. 
\section{The ISRS GN model for arbitrary modulation formats in closed-form}
\label{sec:full_performance_prediction}
In this section, we combine the modulation format correction formula \eqref{eq:cf_modcorr_ISRSGN} with the ISRS GN model contribution in closed form derived in \cite{Semrau_2019_aca}. The result is closed-form formula capable of predicting the total nonlinear performance for arbitrary modulation formats. The proposed formula is once again validated by numerical simulations for the two fiber types with parameters listed in table 1. The formula for the \textit{total} NLI coefficent as in \eqref{eq:total_NLI_coeff} is 
\begin{equation}
\begin{split}
&\eta_{n}\left(f_i\right) \approx \frac{4}{9}\frac{\gamma^2}{B^2_i} \frac{\pi n^{1+\epsilon} }{\phi_{i}\bar{\alpha}\left(2\alpha+\bar{\alpha}\right)}\\
&\cdot\left[\frac{T_i-\alpha^2}{\alpha}\text{asinh}\left(\frac{\phi_{i}B_i^2}{\pi a}\right)+\frac{A^2-T_i}{A}\text{asinh}\left(\frac{\phi_{i}B_i^2}{\pi A}\right)\right]\\
&+\frac{32}{27}\sum_{k=1,k\neq i}^{N_\mathrm{ch}} \left(\frac{P_k}{P_i}\right)^2\frac{\gamma^2}{B_k}\left\{\frac{n+\frac{5}{6}\Phi}{\phi_{i,k}\bar{\alpha}\left(2\alpha+\bar{\alpha}\right)}\right. \\
&\cdot\left[\frac{T_k-\alpha^2}{\alpha}\mathrm{atan}\left(\frac{\phi_{i,k}B_i}{\alpha}\right) +\frac{A^2-T_k}{A}\ \mathrm{atan}\left(\frac{\phi_{i,k}B_i}{A}\right)\right]\\
&\left.+\frac{5}{3} \frac{\Phi\pi \tilde{n} T_k }{\left|\phi\right| B_k^2\alpha^2 A^2}\left[\left(2\left|\Delta f\right|-B_k\right) \log\left(\frac{ 2\left|\Delta f\right| -B_k}{ 2\left|\Delta f\right| +B_k}\right)+ 2B_k\right]\right\},
\label{eq:full_eta_cf}
\end{split}
\end{equation}
with $\phi_i=\frac{3}{2}\pi^2\left(\beta_2+2\pi\beta_3f_i\right)$, $T_k = \left(\alpha+\bar{\alpha}-P_{\text{tot}}C_{\text{r}}f_k\right)^2$, $\Delta f = f_k-f_i$, $\phi_{i,k}=-2\pi^2\left(f_k-f_i\right)\left[\beta_2+\pi\beta_3(f_i+f_k\right]$ and $A=\alpha+\bar{\alpha}$. Eq.  \eqref{eq:full_eta_cf} models the GN model contribution of the SPM and XPM contributions, where the XPM terms are corrected for its modulation format dependence using \eqref{eq:cf_modcorr_ISRSGN}. The reader is referred to \cite{Semrau_2019_aca} for more details about the GN model contribution.
\par 
\ 
The total NLI coefficient as a function of channel frequency with and without ISRS is shown in Fig. \ref{fig:cf_prediction}. The results are shown for SMF and NZDSF. All transmitted channels are entirely modulated with either Gaussian, 16-QAM or 64-QAM symbols. Overall, Eq. \eqref{eq:full_eta_cf} shows very good agreement with the numerical results. In both cases (with and without ISRS), the average mismatch between the closed-form model \eqref{eq:full_eta_cf} and numerical simulation is 0.3 dB and 0.2 dB for SMF and NZDSF, respectively. 
\par 
\ 
The analysis shows that the derived closed-form approximation \eqref{eq:full_eta_cf} is capable of predicting nonlinear performance of ultra-wideband optical transmission systems considering arbitrary modulation formats with sufficient accuracy. A key result in the real-time modelling of next generation ultra-wideband transmission systems.
\section{Conclusion}
A methodology was presented to derive the modulation format correction in closed-form for arbitrary optical transmission systems that are described by a link function. The generic approach is applied to the Gaussian Noise model in the presence of inter-channel stimulated Raman scattering (ISRS GN model) to derive a closed-form formula that predicts the impact of modulation formats on the nonlinear interference in ultra-wideband transmission. The derived formula accounts for arbitrary span numbers, arbitrary launch power distributions, inter-channel stimulated Raman scattering and wavelength dependent dispersion and attenuation. 
\par
\ 
The analytical result was validated by numerical simulations over the entire C+L band (10 THz), with an average prediction error in nonlinear interference of 0.3 dB for standard single-mode fiber (SMF) and 0.2 dB for non-zero dispersion-shifted fiber (NZDSF).
\par 
\ 
The results in this paper are the first analytical investigation of the impact of modulation formats in ultra-wideband optical transmission and the first closed-form formula accounting for it. A significant result for the design of optical communication systems and the real-time performance modeling of ultra-wideband transmission systems. 

\appendices

\section{Proof}
\label{sec:appendix_proof}
In this section, an identity for $\lim_{n \to \infty} \frac{\partial }{\partial n} C_n $ is derived in order to obtain closed-form formula for the asymptotic modulation format correction contribution \eqref{eq:etaa_cf_general}. The asymptotic differential normalization coefficient can be calculated exactly and no approximations are needed. 
\par 
\ 
For notational brevity, the identity is derived for arbitrary parameters $a$ and $b$ which reads 
\begin{equation}
\begin{split}
 C^{'}_{\infty} 
 &=\lim_{n \to \infty}\frac{\partial }{\partial n}\int_{-\infty}^\infty dx \left|\sum_{m=1}^{n}\mathrm{sinc}\left(max\right)e^{jmbx}\right|^2 \\
 &= \frac{\pi}{a^2}\left[\left(b-a\right)\mathrm{ln}\left(\frac{b-a}{a+b}\right)+2a\right].
\label{eq:appendix_statement}
\end{split}
\end{equation}
Executing the derivatve with respect to the span number in \eqref{eq:appendix_statement} yields 
\begin{equation}
\begin{split}
 &C^{'}_{\infty}=\lim_{n \to \infty}\frac{\partial }{\partial n}\int_{-\infty}^\infty dx \left|\sum_{m=1}^{n}\mathrm{sinc}\left(max\right)e^{jmbx}\right|^2 \\
 &=\lim_{n \to \infty}\int_{-\infty}^\infty dx\frac{1}{2a^2nx^2} \sum_{m=1}^{n}\frac{1}{m}\left\{\right.\\
 &\left[\sin\left(mc_2x\right)+\sin\left(mc_1x\right)\right]\left[\sin\left(nc_2x\right)+\sin\left(nc_1x\right)\right]\\
+&\left[\cos\left(mc_2x\right)-\cos\left(mc_1x\right)\right]\left. \left[\cos\left(nc_2x\right)-\cos\left(nc_1x\right)\right]\right\},
\end{split}
\end{equation}
with $c_1=a+b$ and $c_2=a-b$, which are defined for the sake of a concise notation. Using trigonometric identities and the exact integral identity $\int_{-\infty}^{\infty}dx \frac{\sin\left(ax\right)\sin\left(bx\right)}{x^2} = \frac{1}{2}\pi \left(\left|a+b\right|-\left|a-b\right|\right)$ yields 
\begin{equation}
\begin{split}
&C^{'}_{\infty}=\lim_{n \to \infty}\frac{\pi}{2a^2n} \sum_{m=1}^{n}\frac{1}{m}\left\{\right.\\
 &-\left|mc_2-nc_2\right|+\left|mc_1+nc_2\right|+\left.\left|mc_2+nc_1\right|-\left|mc_1-nc_1\right|\right\}.
\label{eq:appendix_123}
\end{split}
\end{equation}
To resolve the absolute value operations in \eqref{eq:appendix_123}, we identify that $b\geq 0$, $a\geq0$ and $b\geq a$, resulting in $c_1\geq 0$ and $c_2 \leq 0$. The conditions are met by default as the minimum feasible channel spacing is $\Delta f \geq \frac{B_k}{2}$ and the result is invariant to the sign of $b$. Applying the conditions results in
\begin{equation}
\begin{split}
C^{'}_{\infty} &=\lim_{n \to \infty}\frac{\pi}{2a^2n} \sum_{m=1}^{n}\frac{1}{m}\\
 &\left\{b\left(m-n\right)+\left(n+m\right)a+\left|mc_1-n\left|c_2\right|\right|\right\}.
\label{eq:appendix_1234}
\end{split}
\end{equation}
Eq. \eqref{eq:appendix_1234} can be written as two distinct series, which are solved separately in the following. The first series is exactly written and solved as
\begin{equation}
\begin{split}
 &\lim_{n \to \infty}\frac{\pi}{2a^2n} \sum_{m=1}^{n}\frac{1}{m}\left\{b\left(m-n\right)+\left(n+m\right)a\right\} \\
 &=\lim_{n \to \infty} \frac{\pi}{2a^2}c_1-\left|c_2\right|\mathrm{Har}\left(n\right), \\
\label{eq:series_1}
\end{split}
\end{equation}
with $\mathrm{Har}\left(n\right)$ being the n'th harmonic number. The second series can be written and exactly solved as
\begin{equation}
\begin{split}
 &\lim_{n \to \infty}\frac{\pi}{2a^2n} \sum_{m=1}^{n}\frac{1}{m}\left|mc_1-n\left|c_2\right|\right| \\
 &=\lim_{n \to \infty}\frac{\pi}{2a^2}\left(\sum_{m=1}^{\floor*{\frac{n\left|c_2\right|}{c_1}}}\frac{\left|c_2\right|}{m}-\frac{c_1}{n} +\sum_{m=\ceil*{\frac{n\left|c_2\right|}{c_1}}}^{n}\frac{c_1}{n}-\frac{\left|c_2\right|}{m}\right) \\
&=\lim_{n \to \infty}\frac{\pi}{2a^2}\left\{\left|c_2\right|\mathrm{Har}\left(\floor*{\frac{n\left|c_2\right|}{c_1}}\right)-\left|c_2\right|\mathrm{Har}\left(n\right)\right. \\
&+\left|c_2\right|\mathrm{Har}\left(\ceil*{\frac{n\left|c_2\right|}{c_1}}\right)\left.+\frac{c_1}{n}\left(n-\ceil*{\frac{n\left|c_2\right|}{c_1}}-\floor*{\frac{n\left|c_2\right|}{c_1}}\right)\right\} \\
&=\lim_{n \to \infty}\frac{\pi}{2a^2} \left[2\left|c_2\right|\mathrm{Har}\left(\frac{n\left|c_2\right|}{c_1}\right)-\left|c_2\right|\mathrm{Har}\left(n\right)+c_1-2\left|c_2\right|\right]
\label{eq:series_2}
\end{split}
\end{equation}
, where it was used that $\lim_{n \to \infty} \ceil*{\frac{nc_2}{c_1}}=\lim_{n \to \infty} \floor*{\frac{nc_2}{c_1}}=\lim_{n \to \infty} \frac{nc_2}{c_1}$. Both series can be further simplified by recalling that $\lim_{n \to \infty} \mathrm{Har}\left(n\right)=\lim_{n \to \infty}\log\left(n\right)+\gamma$, with $\gamma$ being the Euler-Mascheroni constant. 
\par 
\ 
Combining \eqref{eq:series_1} and \eqref{eq:series_2} yields the final result
\begin{equation}
\begin{split}
C^{'}_{\infty}&=\lim_{n \to \infty}\frac{\pi}{a^2}\left\{\left|c_2\right|\mathrm{Har}\left(\frac{n\left|c_2\right|}{c_1}\right)-\left|c_2\right|\mathrm{Har}\left(n\right)+2a\right\} \\
&=\frac{\pi}{a^2}\left[\left|c_2\right|\mathrm{ln}\left(\frac{\left|c_2\right|}{c_1}\right)+2a\right] \\
&=\frac{\pi}{a^2}\left[\left(b-a\right)\mathrm{ln}\left(\frac{b-a}{a+b}\right)+2a\right]
\end{split}
\end{equation}
which proves the identity in \eqref{eq:appendix_statement}. Therefore, Eq.  \eqref{eq:appendix_statement} can be used to write \eqref{eq:etaa_cf_general} in closed-form as \eqref{eq:cf_etaa_corr}, without imposing any additional assumptions.

\section*{Acknowledgment}
Financial support from UK EPSRC through programme grant TRANSNET (EP/R035342/1) and a Doctoral Training Partnership (DTP) studentship to Daniel Semrau is gratefully acknowledged.

\ifCLASSOPTIONcaptionsoff
  \newpage
\fi

\bibliographystyle{IEEEtran}
\bibliography{IEEEabrv,main}

\begin{thebibliography}{10}
\providecommand{\url}[1]{#1}
\csname url@samestyle\endcsname
\providecommand{\newblock}{\relax}
\providecommand{\bibinfo}[2]{#2}
\providecommand{\BIBentrySTDinterwordspacing}{\spaceskip=0pt\relax}
\providecommand{\BIBentryALTinterwordstretchfactor}{4}
\providecommand{\BIBentryALTinterwordspacing}{\spaceskip=\fontdimen2\font plus
\BIBentryALTinterwordstretchfactor\fontdimen3\font minus
  \fontdimen4\font\relax}
\providecommand{\BIBforeignlanguage}[2]{{%
\expandafter\ifx\csname l@#1\endcsname\relax
\typeout{** WARNING: IEEEtran.bst: No hyphenation pattern has been}%
\typeout{** loaded for the language `#1'. Using the pattern for}%
\typeout{** the default language instead.}%
\else
\language=\csname l@#1\endcsname
\fi
#2}}
\providecommand{\BIBdecl}{\relax}
\BIBdecl

\bibitem{Nespola_2014_gvo}
A.~Nespola, S.~Straullu, A.~Carena, G.~Bosco, R.~Cigliutti, V.~Curri,
  P.~Poggiolini, M.~Hirano, Y.~Yamamoto, T.~Sasaki, J.~Bauwelinck, K.~Verheyen,
  and F.~Forghieri, ``{GN-Model} validation over seven fiber types in
  uncompensated {PM}-16{QAM} {Nyquist}-{WDM} links,'' \emph{IEEE Photon.
  Technol. Lett.}, vol.~26, no.~2, pp. 206--209, Jan. 2014.

\bibitem{Cai_2014_tpo}
J.~X. Cai, H.~G. Batshon, H.~Zhang, M.~Mazurczyk, O.~V. Sinkin, D.~G. Foursa,
  and A.~N. Pilipetskii, ``Transmission performance of coded modulation formats
  in a wide range of spectral efficiencies,'' in \emph{OFC 2014}, Mar. 2014,
  pp. 1--3.

\bibitem{Nespola_2015_evo}
A.~Nespola, M.~Huchard, G.~Bosco, A.~Carena, Y.~Jiang, P.~Poggiolini, and
  F.~Forghieri, ``Experimental validation of the {EGN}-model in uncompensated
  optical links,'' in \emph{Optical Fiber Communication Conference
  (OFC)}.\hskip 1em plus 0.5em minus 0.4em\relax Optical Society of America,
  2015, p. Th4D.2.

\bibitem{Galdino_2016_edo}
L.~Galdino, G.~Liga, G.~Saavedra, D.~Ives, R.~Maher, A.~Alvarado, S.~Savory,
  R.~Killey, and P.~Bayvel, ``Experimental demonstration of
  modulation-dependent nonlinear interference in optical fibre communication,''
  in \emph{ECOC 2016; 42nd European Conference on Optical Communication}, Sept
  2016, pp. 1--3.

\bibitem{Saavedra_2017_eio}
G.~Saavedra, M.~Tan, D.~J. Elson, L.~Galdino, D.~Semrau, M.~A. Iqbal, I.~D.
  Phillips, P.~Harper, N.~M. Suibhne, A.~D. Ellis, D.~Lavery, B.~C. Thomsen,
  R.~I. Killey, and P.~Bayvel, ``Experimental investigation of nonlinear signal
  distortions in ultra-wideband transmission systems,'' in \emph{2017 Optical
  Fiber Communications Conference and Exhibition (OFC)}, March 2017, pp. 1--3.

\bibitem{Saavedra_2017_eao}
G.~Saavedra, M.~Tan, D.~J. Elson, L.~Galdino, D.~Semrau, M.~A. Iqbal, I.~D.
  Phillips, P.~Harper, A.~Ellis, B.~C. Thomsen, D.~Lavery, R.~I. Killey, and
  P.~Bayvel, ``Experimental analysis of nonlinear impairments in fibre optic
  transmission systems up to 7.3 {THz},'' \emph{J. Lightw. Technol.}, vol.~35,
  no.~21, pp. 4809--4816, Nov. 2017.

\bibitem{Saavedra_2018_isr}
G.~Saavedra, D.~Semrau, M.~Tan, M.~A. Iqbal, D.~J. Elson, L.~Galdino,
  P.~Harper, R.~I. Killey, and P.~Bayvel, ``Inter-channel stimulated {R}aman
  scattering and its impact in wideband transmission systems,'' in
  \emph{Optical Fiber Communication Conference}.\hskip 1em plus 0.5em minus
  0.4em\relax Optical Society of America, 2018, p. Th1C.3.

\bibitem{Hasegawa_2017_ofd}
T.~Hasegawa, Y.~Yamamoto, and M.~Hirano, ``Optimal fiber design for large
  capacity long haul coherent transmission,'' \emph{Opt. Express}, vol.~25,
  no.~2, pp. 706--712, Jan. 2017.

\bibitem{Semrau_2016_air}
D.~Semrau, T.~Xu, N.~A. Shevchenko, M.~Paskov, A.~Alvarado, R.~I. Killey, and
  P.~Bayvel, ``Achievable information rates estimates in optically amplified
  transmission systems using nonlinearity compensation and probabilistic
  shaping,'' \emph{Optics Lett.}, vol.~42, no.~1, p. 121, Dec. 2016.

\bibitem{Bosco_2011_aro}
G.~Bosco, P.~Poggiolini, A.~Carena, V.~Curri, and F.~Forghieri, ``Analytical
  results on channel capacity in uncompensated optical links with coherent
  detection,'' \emph{Opt. Express}, vol.~19, no.~26, pp. B440--B451, Dec. 2011.

\bibitem{Shevchenko_2016_air}
N.~A. Shevchenko, T.~Xu, D.~Semrau, G.~Saavedra, G.~Liga, M.~Paskov,
  L.~Galdino, A.~Alvarado, R.~I. Killey, and P.~Bayvel, ``Achievable
  information rates estimation for 100-nm {Raman}-amplified optical
  transmission system,'' in \emph{ECOC 2016; 42nd European Conference on
  Optical Communication}, Sept. 2016, pp. 1--3.

\bibitem{Ramamurthy_1999_iot}
B.~{Ramamurthy}, D.~{Datta}, H.~{Feng}, J.~P. {Heritage}, and B.~{Mukherjee},
  ``Impact of transmission impairments on the teletraffic performance of
  wavelength-routed optical networks,'' \emph{J. Lightwave Technol.}, vol.~17,
  no.~10, p. 1713, Oct 1999.

\bibitem{Anagnostopoulos_2007_pli}
V.~Anagnostopoulos, C.~T. Politi, C.~Matrakidis, and A.~Stavdas, ``Physical
  layer impairment aware wavelength routing algorithms based on analytically
  calculated constraints,'' \emph{Optics Communications}, vol. 270, no.~2, pp.
  247--254, Feb. 2007.

\bibitem{Splett_1993_utc}
A.~Splett, C.~Kurtzke, and K.~Petermann, ``Ultimate transmission capacity of
  amplified optical fiber communication systems taking into account fiber
  nonlinearities,'' in \emph{1993 The European Conference on Optical
  Communication (ECOC)}, 1993.

\bibitem{Tang_2002_tcc}
J.~Tang, ``The channel capacity of a multispan {DWDM} system employing
  dispersive nonlinear optical fibers and an ideal coherent optical receiver,''
  \emph{J. Lightw. Technol.}, vol.~20, no.~7, p. 1095, Jul. 2002.

\bibitem{Poggiolini_2012_tgm}
P.~Poggiolini, ``The {GN} model of non-linear propagation in uncompensated
  coherent optical systems,'' \emph{J. Lightw. Technol.}, vol.~30, no.~24, pp.
  3857--3879, Dec. 2012.

\bibitem{Louchet_2003_amf}
H.~Louchet, A.~Hodzic, and K.~Petermann, ``Analytical model for the performance
  evaluation of {DWDM} transmission systems,'' \emph{IEEE Photonics Technology
  Letters}, vol.~15, no.~9, pp. 1219--1221, Sept 2003.

\bibitem{Chen_2010_cef}
X.~Chen and W.~Shieh, ``Closed-form expressions for nonlinear transmission
  performance of densely spaced coherent optical {OFDM} systems,'' \emph{Opt.
  Express}, vol.~18, no.~18, pp. 19\,039--19\,054, Aug. 2010.

\bibitem{Savory_2013_aft}
S.~J. Savory, ``Approximations for the nonlinear self-channel interference of
  channels with rectangular spectra,'' \emph{IEEE Photon. Technol. Lett.},
  vol.~25, no.~10, pp. 961--964, May 2013.

\bibitem{Johannisson_2014_mon}
P.~Johannisson and E.~Agrell, ``Modeling of nonlinear signal distortion in
  fiber-optic networks,'' \emph{J. Lightw. Technol.}, vol.~32, no.~23, pp.
  3942--3950, Dec 2014.

\bibitem{Poggiolini_2015_asa}
P.~Poggiolini, G.~Bosco, A.~Carena, V.~Curri, Y.~Jiang, and F.~Forghieri, ``A
  simple and effective closed-form {GN} model correction formula accounting for
  signal non-{Gaussian} distribution,'' \emph{J. Lightw. Technol.}, vol.~33,
  no.~2, pp. 459--473, Jan. 2015.

\bibitem{Semrau_2017_ace}
D.~Semrau, G.~Saavedra, D.~Lavery, R.~I. Killey, and P.~Bayvel, ``A closed-form
  expression to evaluate nonlinear interference in {R}aman-amplified links,''
  \emph{J. Lightw. Technol.}, vol.~35, no.~19, pp. 4316--4328, Oct 2017.

\bibitem{Chen_2018_g6c}
B.~{Chen}, C.~{Okonkwo}, D.~{Lavery}, and A.~{Alvarado}, ``Geometrically-shaped
  64-point constellations via achievable information rates,'' in \emph{2018
  20th International Conference on Transparent Optical Networks (ICTON)}, July
  2018, pp. 1--4.

\bibitem{Mecozzi_2012_nsl}
A.~Mecozzi and R.-J. Essiambre, ``Nonlinear shannon limit in pseudolinear
  coherent systems,'' \emph{J. Lightw. Technol.}, vol.~30, no.~12, pp.
  2011--2024, Jun. 2012.

\bibitem{Secondini_2012_afc}
M.~Secondini and E.~Forestieri, ``Analytical fiber-optic channel model in the
  presence of cross-phase modulation,'' \emph{IEEE Photon. Technol. Lett.},
  vol.~24, no.~22, pp. 2016--2019, Nov. 2012.

\bibitem{Dar_2013_pon}
R.~Dar, M.~Feder, A.~Mecozzi, and M.~Shtaif, ``Properties of nonlinear noise in
  long, dispersion-uncompensated fiber links,'' \emph{Opt. Express}, vol.~21,
  no.~22, p. 25685, Oct. 2013.

\bibitem{Dar_2014_aon}
------, ``Accumulation of nonlinear interference noise in fiber-optic
  systems,'' \emph{Opt. Express}, vol.~22, no.~12, pp. 14\,199--14\,211, Jun.
  2014.

\bibitem{Jiang_2014_tgm}
Y.~Jiang and P.~Poggiolini, ``The {EGN} model of nonlinear propagation in
  coherent optical transmission systems and its applications,'' \emph{PhD
  Thesis, Politecnico di Torino}, 2014.

\bibitem{Carena_2014_emo}
A.~Carena, G.~Bosco, V.~Curri, Y.~Jiang, P.~Poggiolini, and F.~Forghieri,
  ``{EGN} model of non-linear fiber propagation,'' \emph{Opt. Express},
  vol.~22, no.~13, p. 16335, Jun. 2014.

\bibitem{Ghazisaeidi_2017_ato}
A.~Ghazisaeidi, ``A theory of nonlinear interactions between signal and
  amplified spontaneous emission noise in coherent wavelength division
  multiplexed systems,'' \emph{J. Lightw. Technol.}, vol.~35, no.~23, pp.
  5150--5175, Dec. 2017.

\bibitem{Semrau_17_ard}
D.~Semrau, R.~Killey, and P.~Bayvel, ``Achievable rate degradation of
  ultra-wideband coherent fiber communication systems due to stimulated {Raman}
  scattering,'' \emph{Opt. Express}, vol.~25, no.~12, pp. 13\,024--13\,034,
  Jun. 2017.

\bibitem{Roberts_2018_co}
I.~Roberts, J.~M. Kahn, J.~Harley, and D.~Boertjes, ``Corrections to
  {"}{C}hannel power optimization of {WDM} systems following {G}aussian noise
  nonlinearity model in presence of stimulated {R}aman scattering{"},''
  \emph{Journal of Lightwave Technology}, vol.~36, no.~11, pp. 2309--2309, June
  2018.

\bibitem{Cantono_2018_mti}
M.~Cantono, J.~L. Auge, and V.~Curri, ``Modelling the impact of {SRS} on {NLI}
  generation in commercial equipment: an experimental investigation,'' in
  \emph{Optical Fiber Communication Conference}.\hskip 1em plus 0.5em minus
  0.4em\relax Optical Society of America, 2018, p. M1D.2.

\bibitem{Semrau_2018_tgn}
D.~Semrau, R.~I. Killey, and P.~Bayvel, ``The {G}aussian {N}oise model in the
  presence of inter-channel stimulated {R}aman scattering,'' \emph{J. Lightw.
  Technol.}, vol.~36, no.~14, pp. 3046--3055, July 2018.

\bibitem{Cantono_2018_oti}
M.~Cantono, D.~Pilori, A.~Ferrari, C.~Catanese, J.~Thouras, J.~L. Auge, and
  V.~Curri, ``On the interplay of nonlinear interference generation with
  stimulated {R}aman scattering for {QoT} estimation,'' \emph{J. Lightw.
  Technol.}, pp. 1--1, Aug. 2018.

\bibitem{Semrau_2018_tig}
D.~Semrau, E.~Sillekens, R.~I. Killey, and P.~Bayvel, ``The {ISRS} {GN} model,
  an efficient tool in modeling ultra-wideband transmission in point-to-point
  and network scenarios,'' in \emph{2018 European Conference on Optical
  Communication (ECOC), Tu4G.6}, Sep. 2018, pp. 1--3, pre--print available in
  arxiv:1808.00\,533.

\bibitem{Semrau_2018_aca}
D.~Semrau, R.~I. Killey, and P.~Bayvel, ``A closed-form approximation of the
  {G}aussian {N}oise model in the presence of inter-channel stimulated {R}aman
  scattering,'' \emph{arXiv:1808.07940}, Aug. 2018.

\bibitem{Semrau_2019_aca}
D.~{Semrau}, R.~I. {Killey}, and P.~{Bayvel}, ``A closed-form approximation of
  the gaussian noise model in the presence of inter-channel stimulated raman
  scattering,'' \emph{J. Lightw. Technol.}, pp. 1--1, Jan. 2019.

\bibitem{Poggiolini_2018_agg}
P.~Poggiolini, ``A generalized {GN}-model closed-form formula,''
  \emph{arXiv:1810.06545v2}, Sep. 2018.

\bibitem{Fehenberger_2016_ops}
T.~Fehenberger, A.~Alvarado, G.~B\"{o}cherer, and N.~Hanik, ``On probabilistic
  shaping of quadrature amplitude modulation for the nonlinear fiber channel,''
  \emph{J. Lightw. Technol.}, vol.~34, no.~21, pp. 5063--5073, Nov 2016.

\bibitem{Kschischang_1993_ons}
F.~R. {Kschischang} and S.~{Pasupathy}, ``Optimal nonuniform signaling for
  gaussian channels,'' \emph{IEEE Transactions on Information Theory}, vol.~39,
  no.~3, pp. 913--929, May 1993.

\end{thebibliography}

\begin{IEEEbiographynophoto}{Daniel Semrau}
(S’16) received the B.Sc. degree in electrical engineering from the Technical University of Berlin, Berlin, Germany, in 2013, the M.Sc. degree in photonic networks engineering from Scuola Superiore Sant’Anna, Pisa, Italy, and Aston University, Birmingham, U.K., in 2015. In 2015, he joined the Optical Networks Group, University College London, U.K., where he is currently working toward the Ph.D. degree. In 2018, Daniel was presented with the Graduate Student Fellowship award of the IEEE Photonics Society. His research interests are mainly focused on channel modeling, nonlinear compensation techniques, and ultra-wideband transmission coherent optical communications.
\end{IEEEbiographynophoto}

\begin{IEEEbiographynophoto}{Eric Sillekens}
(S’16) received his BSc and MSc in electrical engineering from the Eindhoven University of Technology in 2012 and 2015 respectively, with his research focussed on advanced coded modulation for optical fibre transmission systems. He is currently a PhD research student in the optical networks group at Univesity College London (UCL) and is supervised by Dr. R. Killey. He is working to holistically optimise long-haul fibre transmission systems, where his interest is in coded modulation and machine learning. He has designed modulation formats with a trade off between fibre nonlinearity and shaping gain.
\end{IEEEbiographynophoto}

\begin{IEEEbiographynophoto}{Robert I. Killey}
(SM’17) received the B.Eng. degree in electronic and communications engineering from the University of Bristol, Bristol, U.K., in 1992, the M.Sc. degree from University College London (UCL), London, U.K., in 1994, and the D.Phil. degree from the University of Oxford, Oxford, U.K., in 1998. He is currently an associate professor with the Optical Networks Group with UCL. His research interests include nonlinear fiber effects in WDM transmission, advanced modulation formats, and digital signal processing for optical communications. He has participated in many European projects, including ePhoton/ONe, Nobel, BONE and ASTRON, and national projects. He is currently a Principal Investigator in the EPSRC funded UNLOC project. He was with the technical program committees of many international conferences including European Conference on Optical Communication, Optical Fiber Communication Conference ACP, and OECC. He was an Associate Editor of the IEEE/OSA Journal of Optical Communications and Networking and is currently an Associate Editor of the Journal of Lightwave Technology.
\end{IEEEbiographynophoto}

\begin{IEEEbiographynophoto}{Polina Bayvel}
(F'10) received the B.Sc. (Eng.) and Ph.D. degrees in electronic and electrical engineering from UCL (University of London), in 1986 and 1990, respectively. 
In 1990, she was with the Fiber Optics Laboratory, General Physics Institute,
Moscow, Russian Academy of Sciences, under the Royal Society Postdoctoral
Exchange Fellowship. She was a Principal Systems Engineer with STC Sub-
marine Systems, Ltd., London, U.K., and Nortel Networks (Harlow, U.K., and
Ottawa, ON, Canada), where she was involved in the design and planning of
optical fibre transmission networks. During 1994–2004, she held a Royal
Society University Research Fellowship at University College London (UCL),
London, U.K., where she became a Chair in Optical Communications and Net-
works. She is currently the Head of the Optical Networks Group, UCL, which
she set up in 1994. She has authored or coauthored more than 300 refereed jour-
nal  and  conference  papers.  Her  research  interests  include  wavelength-routed
optical networks, high-speed optical transmission, and the study and mitigation
of fibre nonlinearities. She is a Fellow of the Royal Academy of Engineering, IEEE,
the Optical Society of America and the U.K. Institute of Physics. She is Honorary Fellow of the Institution of Engineering and Technology (FIET). She was a recipient the Royal Society Wolfson
Research Merit Award (2007–2012), the 2013 IEEE Photonics Society Engi-
neering  Achievement  Award, the  2014  Royal  Society  Clifford  Patterson
Prize Lecture and Medal and 2015 Royal Academy of Engineering Colin Campbell Mitchell Award. She leads the UK EPSRC Programme TRANSNET (2018-2024).
\end{IEEEbiographynophoto}

\end{document}